\documentclass[10pt,conference]{IEEEtran}
\usepackage{geometry}
\usepackage{amsmath} 
\allowdisplaybreaks
\usepackage{algorithm}  
\usepackage{algpseudocode,algorithmicx}
\usepackage{slashed} 
\usepackage{wasysym}
\usepackage{bm}
\usepackage{mathrsfs}
\usepackage{graphicx}  
\usepackage{breqn}
\usepackage{cite}
\usepackage{enumitem}
\usepackage{amsmath}
\usepackage{array}
\usepackage{stfloats}

\begin{document}

\title{Re-DPoctor: Real-time health data releasing with w-day differential privacy}

\author{\IEEEauthorblockN{Jiajun Zhang\IEEEauthorrefmark{1},
Xiaohui Liang\IEEEauthorrefmark{2},
Zhikun Zhang\IEEEauthorrefmark{3},
Shibo He\IEEEauthorrefmark{3} and
Zhiguo Shi\IEEEauthorrefmark{1}
%Eldon Tyrell\IEEEauthorrefmark{4}
}
\IEEEauthorblockA{\IEEEauthorrefmark{1}College of Information Science\&Electronic Engineering,
Zhejiang University, China Email:\{justinzhang,~shizg\}@zju.edu.com}
\IEEEauthorblockA{\IEEEauthorrefmark{2}Department of Computer Science, University of Massachusetts Boston，
	USA Email: xiaohui.liang@umb.edu }
\IEEEauthorblockA{\IEEEauthorrefmark{3}College of Control Science \& Engineering,
	Zhejiang University, China Email:\{zhangzhk,~s18he\}@zju.edu.com}
}

\maketitle

\begin{abstract}
Wearable devices enable users to collect health data and share them with healthcare providers for improved health service. Since health data contain privacy-sensitive information, unprotected data release system may result in privacy leakage problem. Most of the existing work use differential privacy for private data release. However, they have limitations in healthcare scenarios because they do not consider the unique features of health data being collected from wearables, such as continuous real-time collection and pattern preservation. In this paper, we propose Re-DPoctor, a real-time health data releasing scheme with $w$-day differential privacy where the privacy of health data collected from any consecutive $w$ days is preserved. We improve utility by using a specially-designed partition algorithm to protect the health data patterns. Meanwhile, we improve privacy preservation by applying newly proposed adaptive sampling technique and budget allocation method. We prove that Re-DPoctor satisfies $w$-day differential privacy. Experiments on real health data demonstrate that our method achieves better utility with strong privacy guarantee than existing state-of-the-art methods.
\end{abstract}

\IEEEpeerreviewmaketitle

\section{Introduction}

The proliferation of wearable devices, such as Fitbit and Apple Watch, enables the continuous collection of personal health data including heart rate, walking steps, and sleep condition. The personal health data can be a good indication for users to keep track of their ﬁtness and can be further shared with healthcare providers for various purposes. For example, users could share data with an insurance company for a lower premium, and ﬁtness advisor for a better health plan. In these cases, users prefer to share minimum amount of information to healthcare providers. From \cite{Lane2010Balancing}, the disclosure of unnecessary health data may result in severe privacy violations. We consider a scenario where a healthcare provider requires a user to provide the health data collected during the next two weeks. The user needs to consider two factors, i) utility, the disclosed data must be useful; ii) privacy, the disclosure must consume less than a privacy budget.

Health data collected from wearable devices has following unique properties. First, it contains signiﬁcant health patterns, which may imply health conditions. The patterns need to be reserved in the privacy protection algorithm. Second, health data is generated continuously. The usefulness of data varies from day to day. Generally, when the data is not useful, the data does not need to be disclosed. On the other hand, if the data is useful, the data need to be disclosed with a privacy constraint. Given a privacy budget for two weeks, for example, the budget should be adaptively arranged on a daily basis. As such, the utility of the disclosed data can be maximized while the privacy goal is achieved.

Differential Privacy \cite{Dwork2006Differential}, proposed by DWork, is a popular paradigm to provide privacy in the data release. A common way to achieve differential privacy is to perturb data with noise \cite{Dwork2008Differential,Dwork2004Privacy}. Most existing literatures has mainly focused on the one-time release of static data \cite{Dwork2006Calibrating,duan2017distributed,Xiao2011iReduct,Xu2013Differentially,Kellaris2013Practical}. However, in health releasing scenario, data has to be collected and released continuounsly due to the power limit of wearable devices. Several studies\cite{kellaris2014differentially,fan2014adaptive,Li2015Differentially,he2016full} have been focused on real-time data releasing with differential privacy guarantee.  In \cite{Wang2016RescueDP}, Wang et al. proposed a scheme achieving $w$-event privacy. However, their schemes have limitations. Its decision on data usefulness only depends on the data dynamics and ignore the health condition of the user. Thus it does not fit in our case.

In this paper, we propose Re-DPoctor for Real-time e-doctor health data releasing with differential privacy to solve our problem. The contributions of this paper can be summarized as follows.

$\bullet$ We proposed a practical releasing scheme Re-DPoctor which guarantees $w$-day privacy, a new privacy level deﬁnition in the continuous data stream. Its key modules include adaptive sampling, adaptive budget allocation, DP-Partition, perturbation, feature extraction and ﬁltering.
 	
$\bullet$ The design of Re-DPoctor achieves better accuracy and privacy level. It uses partition algorithm to protect health pattern to improve the accuracy while using adaptive sampling and budget allocation algorithm which takes health condition and data dynamic into account to improve privacy level.
 	
$\bullet$ We prove that our scheme satisﬁes $w$-day privacy and do experiments on real collected wearable device data. Compared to others, we have better results on utility and privacy guarantee.

\section{Preliminaries}

\subsection{Differential Privacy}
A mechanism which satisfies Differential Privacy should guarantee that the query result remains approximately the same if a single record is added or deleted.

\emph{\textbf{Definition 1 (Differential Privacy\cite{Dwork2006Differential})}:} A randomized mechanism $\mathcal{M}$ gives $\epsilon$-differential privacy if for all data sets $D_1$ and $D_2$ differing on at most one, and all   $O \subseteq Range(\mathcal{M})$,
\begin{equation}\label{1}
{\rm Pr}[\mathcal{M}(D_1)\in O]\leq {\rm exp}(\epsilon)\cdot {\rm Pr}[\mathcal{M}(D_2)\in O]
\end{equation}
$\epsilon$ is the privacy budget. A smaller $\epsilon$ means more noise and stronger privacy level.

Laplace mechanism is the most common one to guarantee $\epsilon$-differential privacy.

\emph{\textbf{Theorem 1 (Laplace Mechanism\cite{Dwork2006Calibrating})}:} For any function $f:\mathcal{D} \rightarrow \mathcal{R}^{d}$, the Laplace Mechanism $f$ for any dataset $D\in\mathcal{D}$  
\begin{equation}\label{3}
\mathcal{M}(D) = f(D) + Lap(\frac{\Delta(f)}{\epsilon})
\end{equation}
satisfies $\epsilon$-differential privacy. Here, $\Delta(f)$ is sensitivity defined in\cite{Dwork2006Calibrating} and $\epsilon$ represents the privacy level.

%\emph{\textbf{Theorem 2 (Sequential composition\cite{Mcsherry2009Privacy})}:}  $\mathcal{M}_1$,\dots,$\mathcal{M}_k$ is $k$ algorithms that independently satisfy $\epsilon_1-DP$,\dots,$\epsilon_k-DP$, respectively. Publishing the outputs of all $k$ algorithms on an input $D$ satisfies $(\sum_{i=1}^{k} \epsilon_i)-DP$.

\subsection{$w$-day Privacy}
$w$-day $\epsilon$-differential privacy is a concept improved from \cite{kellaris2014differentially}, which is a new way to define privacy level over infinite stream information. It guarantees that for any successive events happened in a window of $w$ days; the privacy leakage level is no more than $\epsilon$.

We model the data stream as an infinite stream tuple $S=(D_1,D_2,...)$, where $S[i]$ is the $i^{th}$ element of $S$, i.e. $D_i$. The stream prefix of $S$ at $t$ represents as $S_t=(D_1,D_2,...,D_t)$. 

\emph{\textbf{Definition 2 ($w$-neighboring)}:} Let $w$ to be a positive integer. Two stream prefixs $S_t$,$S_t'$ are $w$-neighboring, if

\begin{enumerate}
	\item for each pair $S_t[i] \neq S_t'[i]$ with $i\in [t]$, it holds that $S_t[i]$,$S_t'[i]$ are neighboring (e.g.$S_t[i]$,$S_t'[i]$ have at most one row different);
	\item for each $S_t[i_1]$,$S_t[i_2]$,$S_t'[i_1]$,$S_t'[i_2]$ with $i_1 < i_2$,$S_t[i_1]\neq S_t'[i_1]$and $S_t[i_2]\neq S_t'[i_2]$, it holds that $i_2-i_1+1\leq w$.
\end{enumerate}

\emph{\textbf{Definition 3 ($w$-day Privacy)}:} Let $\mathcal{M}$ be a mechanism that takes as input a stream prefix of arbitrary size. Let $\mathcal{O}=Range(\mathcal{M})$ be the set of all possible outputs of $\mathcal{M}$. Then we call that $\mathcal{M}$ satisfies $w$-day $\epsilon$-differential privacy if for all sets $O \subseteq \mathcal{O}$, all $w$-neighboring stream prefixes $S_t[i]$,$S_t'[i]$ , and all $t$, it holds that 
\begin{equation}\label{4}
{\rm Pr}[\mathcal{M}(S_t)\in O]\leq {\rm exp}(\epsilon)\cdot {\rm Pr}[\mathcal{M}(S_t')\in O]
\end{equation}

\emph{\textbf{Theorem 2\cite{kellaris2014differentially}}:} Let $\mathcal{M}$ be a mechanism that takes as input stream prefix $S_t$, where $S_t[i]$ = $D_i \in \mathcal{D}$, and outputs a transcript $o = (o_1,...,o_t) \in \mathcal{O}$. Suppose that we can decompose $\mathcal{M}$ into $t$ mechanisms $\mathcal{M}_1$, . . . , $\mathcal{M}_t$, such that $\mathcal{M}_i(D_i)$ = $o_i$, Let $\mathcal{M}_i$ be $\epsilon_i$-differential private for some $\epsilon_i$. Then, $\mathcal{M}$ will satisfy $w$-differential privacy if 
\begin{equation}\label{5}
\forall i \in[t],\sum\limits_{k=i-w+1}^{i} \epsilon_k \leq \epsilon    
\end{equation}

It means we could view $\epsilon$ as the whole privacy budget in a $w$-day sliding window and any budget falls out of the window could be recycled and reused.

\section{Re-DPoctor: Real-time health data releasing with w-day privacy}

Consider the scenario where the user has a wearable device to monitor his health data. Also, there exists an E-doctor that the wearable tracking device would release heart rate data to the server in hospital from time to time. When the user goes to the hospital, the doctor can pull out the data and do the analysis. However, the dilemma is, how could we design the health histogram releasing mechanism to only release useful data for diagnosing needs while maintaining the privacy? One common way is to perturb the data with noise. But applying unifying noise to the original data will cause the decreasing precision of histogram. Besides, there are many patterns in the original histogram that could be buried in too much noise. The solution is to design a mechanism that could preserve the desired patterns and protect the privacy.

In this section, we present a real-time health data releasing with $w$-day differential privacy. Figure 1 shows an overview of the proposed scheme, which contains six modules: Partitioning, Perturbation, Feature Extraction, Adaptive Sampling, Adaptive Budget allocation, Filtering. 

\begin{figure}[htb]
			\centerline{\includegraphics[height=4cm,clip]{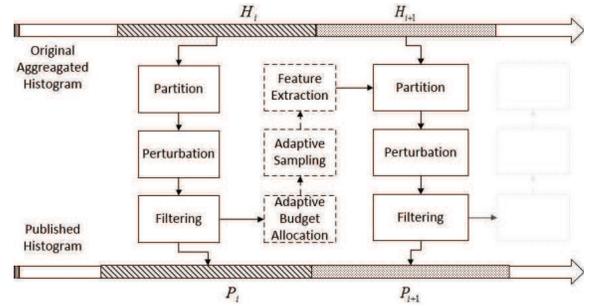}}	
			\caption{Overview of \emph{Real-time e-doctor histogram releasing with differential privacy}}		
\end{figure}

Firstly, \emph{adaptive sampling mechanism} adjusts the sampling rate based on data dynamics and health condition, which perturbs histograms at sampling day and approximate the non-sampled day with perturbed histograms at last sampling day. Then \emph{budget allocation mechanism} dynamically allocates the privacy budget $\epsilon$ at sampling days. The first two steps make sure the non-sampled points can be approximated without any budget allocation. Thus, given a fixed $\epsilon$ more precious privacy budget can be allocated to the histogram needed to be released and reduce the errors caused by Laplace noise and improve overall accuracy. Then, the \emph{DP-Partitioning mechanism} could preserve desired patterns for health diagnose. Then Laplace mechanism is used to perturb the partitioned histogram. At last,\emph{filtering mechanism} helps to improve the accuracy of the released data.  

The followings are the main components of the proposed scheme in details.

\subsection{Adaptive Sampling}
When a user publishes all the histograms at every day, it will introduce large noise and affect the utility of the released histograms. Here comes the seemingly non-negotiable tradeoff between the accuracy and privacy of the histogram releasing. Thus, sampling will be a great method to deal with such a dilemma that we sample the important histogram at certain selected days and leave the non-sampled ones to be approximated. Since the non-sampled histograms do not cost any privacy budget, the selected one can be allocated more budget and improve their accuracy.

Several earlier researchers have proposed methods to adjust sampling rate but didn't fit in our scenario of health data. DSAT\cite{Li2015Differentially} failed to apply in health data because it uses a fixed sampling rate which is unrealistic in real-time health monitoring. Another approach by Wang \cite{Wang2016RescueDP} fails to fit in health monitoring because it ignores the health condition of the user as a dynamic factor which could affect the sampling rate.

In this paper, we proposed a new adaptive sampling mechanism, which takes the current health condition, histogram dynamics, and remaining budget into consideration. Suppose the current sample day is $t_i$ and the last sample day is $t_{i-1}$. The heart rate records are $d_{t_i},d_{t_{i-1}}$ respectively. We use Pearson correlation coefficient as the feedback error:

\begin{equation}\label{Pearson}
E_{t_i}=\rho_{d_{t_i},d_{t_{i-1}}}=\frac{Cov(d_{t_i},d_{t_{i-1}})}{\sigma_{d_{t_i}}\sigma_{d_{t_{i-1}}}}
\end{equation}

Here we choose to use the released histogram instead of the raw histogram to protect the privacy. It may introduce a little error which is relatively small compared to the privacy it provides.

The PID error is defined as:
\begin{equation}\label{PIDerror}
u_{t_i}=\theta_P\times e_{t_i}+\theta_I\times \frac{\sum_{o=t_i-w+1}^{t_i}e_{o}}{w}+\theta_D\times \frac{e_{t_i}}{t_i-t_{i-1}}
\end{equation}
where the $\theta_P$,$\theta_I$,$\theta_D$  are the proportional gain, the integral gain, and the derivative gain.

\textbf{Proportional term:} The first term is proportional to the current error $e_{t_i}=\frac{|E_{t_i}-\delta|}{\delta}$
where $E_{t_i}$ is the feedback error, and the parameter $\delta$ is the set point. We set $\delta$ as 5\% experiments as the maximum tolerance of the feedback error.

\textbf{Integral term:} The second term stands for the accumulation of past error $\theta_I\times \frac{\sum_{o=t_i-m+1}^{t_i}e_{o}}{w}$ where $\theta_I$is the integral gain and the $m$ is how many samples are taken into account. 

\textbf{Derivative term:} The third term $\frac{e_{t_i}}{t_i-t_{i-1}}$ just determines the slope of error over time and predicts the future error.

Intuitively, the sampling interval should be small if user's health condition changes rapidly. However, if the remaining budget is small, sampling at the next day will introduce a high perturbation error. A more reasonable choice is to use a relatively large sampling interval so that previously allocated budget could be recycled and to approximate the histogram with the previous publication.

Besides histogram dynamics and remaining privacy budget, another factor we need to consider is the health condition of the user. Imagine two users have same histogram dynamics and remaining privacy budget but one in sick condition and another one in good health. Applying same sampling method are not applicable because the sick user apparently needs more concerns and needs to release histograms more frequently than the healthy one. One rule for health data releasing is that we should never sacrifice the user's health for privacy. We use $c_{t_i}$ to denote user's health condition which can get from the feature extraction module.

Combined all the three factors, the next sampling rate is defined as below:
\begin{equation}\label{interval}
I_{t_i}=max\{1,I_{t_{i-1}}+\eta(1-(\frac{u_{t_i}}{\lambda})^2),I_{t{i-1}}+\eta(1-(\frac{c_i}{\lambda})^2)\}
\end{equation}
where $I_{t_i}$ and $I_{t_{i-1}}$ is the next and last sampling interval respectively. And $\lambda_r=1/\epsilon_r$ is the scale of Laplace noise where $\epsilon_r$ is the remaining budget. $\eta$ is the scale factor to adjust the sampling interval. Consequently, the sampling interval will increase when the $u<\lambda$ or $c<\lambda$ and decrease otherwise.

\subsection{Adaptive Budget Allocation}
The definition of the $w$-day privacy requires the total budgets within the sliding window of $w$ equals a certain value $\epsilon$. 

For the $i_{th}$ sampling day , firstly, we have to calculate the remaining budget in the window $\epsilon_r=\epsilon-\sum_{j=t_i-w+1}^{t_i-1}\epsilon_j$. Note that if $\epsilon_j$is not a sampling day, then it equals zero. Then, inspired by RescueDP, we allocate the remaining budget based on the sampling interval. When the sampling interval is small, it can be inferred that the histogram changes rapidly or the user is the sick condition. Moreover, we can infer there will be a large number of sampling points in the $w$ time windows. Then, we allocate a small portion of the remaining privacy budget to the coming sampling point so that there will be more privacy left for future use. Fortunately, natural logarithm could quantify such a relationship. Define the portion as:
\begin{equation}\label{portion}
p =min({\rm ln}(\phi \cdot I+1),p_{max})
\end{equation}
where the $\phi$ is the scale factor to adjust the budget portion and the $p_{max}$ limits the maximum value of a portion. So the allocated budget portion will increase as the sampling interval increase. Meanwhile, it slows down when the interval is large enough. Finally, we calculate the budget simply by applying the portion to the remaining budget as $\epsilon_i={\rm min}(p \cdot \epsilon_r,\epsilon_{max})$,where the $\epsilon_{max}$ limits the maximum value of budget because excessive privacy budget could achieve little improvement to the utility of histogram.

\subsection{Partitioning}
Health data histogram is different from other ordinary histograms. Without suitable partition, health data histogram could easily lose their important features or patterns, which are crucial for diagnoses, during aggregation and randomization. The main goal is to design an algorithm to preserve the desired pattern of heart rate in releasing the histogram. We use partition algorithm to protect certain patterns. In our case, we mainly focus on two patterns: small but rapid change and slow but large change.

Before partition, the database records will be aggregated into data bins on a 10 minutes basis. Then the bins will be partitioned into the set of buckets based on the value, the structure and the threshold of the original bins database. Since the buckets structure may reveal information, and one could infer private information in the database due to the small changes in the database. To prevent such privacy leakage, we decide to use part of the privacy allocated for the $i_th$ sampling point to protect the threshold of the partition. Here we use a constant $q$ as the scale to denote the portion of privacy budget for partition.

The algorithm of partitioning with differential privacy is in Algorithm 2. Before the start of the algorithm, several variables need to be declared: Variables $d_i,b_j$ are the value of $i_{th}$ bin of histogram database $D$ and the $j_{th}$ bucket, respectively. Integers $i,j,size$ are the indexes of the current bin and the current bucket and the size of the current bucket,respectively. $last$ holds the value of last bin. The $Min, Max$ indicates the maximum and minimum value of current bucket. And three thresholds which are learned from public information and are set based on user setup:
\begin{itemize}
	\item $T_D$: the maximum difference between the maximum and minimum value in one bucket, accords to slow but large change 
	\item $T_R$: the maximum instant change of heart rate between adjacent bins. Normally, this threshold is smaller than $T_D$ because the change between two adjacent bins may actually be smaller than $T_D$, but since it happened in a very small period of time, it must be preserved. It accords to rapid change. 
	\item $T_S$: the maximum size of each bucket in case of the oversize of a bucket.
\end{itemize}

Due to the privacy requirement of the partition algorithm, we add Laplace noises $Z, Z'$ to $T_D$ and $T_R$ threshold parameters and get $\hat{T_D}$ and $\hat{T_R}$.

\begin{algorithm}
	\caption{Differential-private partition Algorithm}   
	\label{DP-partition}   
	\begin{algorithmic}[1] %这个1 表示每一行都显示数字  
		\Require $D_{t_i}$, $T_D$, $T_R$, $T_L$, $q\cdot \epsilon_i$;
		\Ensure  histogram buckets $B$;\\
		\textbf{Initialization}: Set $size=0;i=1;j=1;B=\emptyset;$
		\State $\hat{T_D}=T_D+Z$, $\hat{T_R}=T_R+Z'$
		\Comment{$Z,Z'\AC Lap((q\cdot \epsilon_i))$}
		\State $b_j\leftarrow d_i$; $Min=Max=current=d_i;size++;i++$;
		\While{$i\leq length(D)$} 
		\If {$current\neq Null$ and $|current-d_i|>\hat{T_R}$}
		\If {$b_{j-1}.length>1$} \\
		\Comment{Last bucket is not a single bin bucket}
		\State $last = B.pop()$; $b_j=last.pop()$;
		\State $B\leftarrow last$;$B\leftarrow b_j$;$j++$; $b_j\leftarrow d_i$;$B\leftarrow b_j$;
		\State $j++$;$current=x$;$i++$;$size=0$;
		\Else \Comment{Last bucket is a single bin bucket}
		\State $b_j\leftarrow x$;$B\leftarrow b_j$;
		\State $j++$,$current=d_i$;$i++$;$size=0$;
		\EndIf
		\ElsIf{$size==1$}
		\State $B\leftarrow b_j$;$j++$;$b_j\leftarrow d_i$;$j++$;
		\State $current=d_i$;$size=0$;$i++$; 
		\ElsIf{$size\geq 1$}
		\State $last = b_j.pop()$;$B\leftarrow b_j$;$j++$; $b_j\leftarrow last$;
		\State $B\leftarrow b_j$;$j++$; $b_j\leftarrow d_i$;$B\leftarrow b_j$;$j++$;
		\State $current=x$;$i++$;$size=0$;
		\EndIf
		\State $Max=max(Max,d_i);Min=min(Min,d_i);$
		\If{$\left| Max-Min\right| \leq \hat{T_D}$ and $size\leq T_S$} 
		\State $b_j\leftarrow d_i;current=d_i;size++;j++;$
		\Else
		\State $B\leftarrow b_j;current=d_i;size=0;j++;$
		\EndIf
		\EndWhile\\
		\textbf{return} $B$
	\end{algorithmic}  
\end{algorithm} 

The partition process could be easily understood. In the beginning, it put the first bin into the first bucket and move to next bin. Then the algorithm checks all the threshold requirement, if they are all met then the current bin will be put into the same bucket. Otherwise, a new bucket will be created. The first checked threshold is $T_R$ due to its smaller value. If the threshold is breached, two single bin buckets need to be created, each containing the adjacent sudden change bins so that their values won't be averaged later. Based on the size of the current bucket, three cases are considered. Moreover, the second and third threshold will be tested and either the new bucket will be created, or the current bucket will be enlarged. 

\subsection{Perturbation}
The results from the previous step buckets then will be randomized by simply adding noise which following Laplace distribution at each sampling point.

After suitable partition, we firstly have to average the bins in the same bucket first. Then, we just add Laplace noise to the average value of bins of every bucket. Suppose the minimum possible change in the query result from neighborhood databases is $\alpha$ and the remaining portion for randomization is $(1-q)\cdot \epsilon_i$. So Laplace noise for $i_{th}$ sampling day will be
\begin{equation}\label{key}
v'_j=v_j+Lap(\frac{\alpha}{(1-q)\cdot \epsilon_i})
\end{equation}
where $v$ is the average value of bucket $j$.

\subsection{Filtering}
In order to eliminate the error introduced by using released data in adaptive sampling and budget allocation mechanism, we use Particle filter improve the accuracy of releasing histogram by estimating the perturbed histogram. We chose Particle filter instead of Kalman filter because in \cite{fan2014adaptive}, it is proved that although the Particle filter cost much more time and has greater complexity, it achieves more accuracy. Moreover, when comes to protect the health data, accuracy weighs better importance than algorithm complexity. In the final releasing histogram $p_i$ at the $i$, it releases posterior estimates of particle filter at sampling points and prior estimates at non-sampling points. Due to the space limit, we omit the details of filtering. Please refer to\cite{fan2014adaptive} for details.

\subsection{Feature Extraction}
Then we need to level the health condition by extracting features from the released histograms. Here we adopt the simplest model just for explanation and focus on four features of four typical rhythms for potential heart disease:
$h_r$: the number of time when the user's heart rate has a rapid increase or decrease in a short period, which could be explained as the signal of heart-attack;
$h_g$: the number of time when the user's heart rate has a great increase or decrease in a long time, which could be explained as the signal of palpitation;
$h_h$: the time when the user's heart rate keeps above maximum threshold, which could be explained as the signal of angina;
$h_l$: the time when the user's heart rate keeps below minimum threshold, which could be explained as the signal of sinus bradycardia；

Then we define the health condition $c_i$ at  $i$ as:
\begin{equation}\label{health}
c_i= max\{\frac{1}{4}(\frac{h_r}{n_r}+\frac{h_g}{n_g}+\frac{h_h}{n_h}+\frac{h_l}{n_l}),1\}
\end{equation}
where $n_r,n_g,n_h,n_l$ are the standard tolerant values from medical references. So the calculated health condition $c_i$ could be used in the adaptive sampling mechanisms. Since the feature extraction is based on the released histogram, so it does not cost any privacy budget, either.

\subsection{Privacy Analysis}

\emph{\textbf{Theorem 3}:} Partitioning process satisfies $q\cdot \epsilon_i$-differential privacy at the  $i$.

\emph{Proof:} Let the $d_0$, $d_1$ be the neighboring databases and the $\mathcal{M}(d_0),\mathcal{M}(d_1)$ be the output. To prove partition process is $q\cdot \epsilon_i$-differential private, we need to prove: $Pr(\mathcal{M}(d_0)=B)\leq e_{q\cdot \epsilon_i} \times Pr(\mathcal{M}(d_1)=B)$. Suppose the maximum difference in the value of bins in two neigboring databases is bounded by $\alpha$. For each bucket, we have to meet the bound $Max_j-Min_j<\hat{T_D}$ and $|current-x_i|<\hat{T_R}$. And according to the sequential composition property of DP, taking $q\cdot \epsilon_i=\epsilon_1+\epsilon_2$. So the inequality can be transformed into: 
\begin{multline*}
\frac{Pr(\mathcal{M}(d_0)=B)}{Pr(\mathcal{M}(d_1)=B)}\leq e^{q\cdot \epsilon_i}\\
\Leftrightarrow \mathcal{X}=(\frac{\prod\limits_{b_i\in d_0}Pr(Max_{j0}-Min_{j0}<\hat{T_D})}{\prod\limits_{b_i\in d_0}Pr(Max_{j0}-Min_{j0}<\hat{T_D})}\leq e^{\epsilon_1})\\
\times( \frac{\prod\limits_{b_i\in d_0}Pr(|current-x_{j0}|<\hat{T_R})}{\prod\limits_{b_i\in d_0}Pr(|current-x_{j1}|<\hat{T_R})}\leq e^{\epsilon_2})
\end{multline*} 

We try to solve the inequalities separately in order to find the required Laplace distribution. Suppose the changed record between the neighbouring databases falls into the bucket $b_j$. 

For the first inequality, the changed record may effect $Max_{j0}$ and $Min_{j0}$ or an ordinary bin's count of $b_j$. If the changed value only affects ordinary bins. Clearly, $\mathcal{X}_1=1<e^{\epsilon_1}$. If the changed value effects either $Max_{j0}$ or $Max_{j1}$, we need to find the suitable Laplace scale($b=s/\epsilon_1$) in order to have this change tolerated. Suppose the $Max_{j0}$ and $Min_{j1}$ are changed by $\alpha$. Take $Z\sim Lap(s/\epsilon_1),t=Max_{j0}-Min_{j1}$ and $u=t-T_D$. Here we only consider the change of $Max_{j0}$.

When $Max{j0}=Max_{j0}+\alpha$:
\begin{equation*}\label{case1}
\mathcal{X}_1=\frac{t+\alpha<\hat{T_D}}{t<\hat{T_D}}=\frac{Z>u+\alpha}{Z>u}<1\leq e^{\epsilon_1}
\end{equation*}

When $Max{j0}=Max_{j0}-\alpha$:
\begin{equation*}\label{case1}
\mathcal{X}_1=\frac{t-\alpha<\hat{T_D}}{t<\hat{T_D}}=\frac{Z>u-\alpha}{Z>u}=\frac{\int_{u-\alpha}^{+\infty}f_z(z)dz}{\int_{u}^{+\infty}f_z(z)dz}\leq e^{\epsilon_1}
\end{equation*}

And we discuss the above inequation in three cases:
\begin{itemize}
	\item $u\geq\alpha$:
	$\mathcal{X}_1=e^{\alpha\epsilon_1/s}\Rightarrow \alpha\epsilon_1/s\leq \epsilon_1 \Rightarrow s \geq \alpha$
	\item
	$0<u<\alpha$:
	\begin{equation*}
	\mathcal{X}_1=\frac{1/2+\int_{u-\alpha}^{0}f_z(z)dz}{\int_{u}^{+\infty}f_z(z)dz}=\frac{2-e^{\frac{u-\alpha}{b}}}{e^{\frac{-u}{b}}}\leq e^{\epsilon_1}
	\end{equation*} 
	
	Let $v=e^{u/b}$,then $\mathcal{X}_1=2v-e^{\frac{-\alpha}{b}}v_2\leq e^\epsilon_1 \Rightarrow s\geq \alpha$
	
	\item $u\leq 0$:
	\begin{multline*}
	\mathcal{X}_1=\frac{1/2+\int_{u-\alpha}^{0}f_z(z)dz}{1/2+\int_{u}^{0}f_z(z)dz}=\frac{2-e^{\frac{u-\alpha}{b}}}{2-e^{\frac{u}{b}}}\leq e^{\epsilon_1}\\\Leftrightarrow e^{\epsilon_1}(e^{u\epsilon_1})^{1/s}-[e^{(u-\alpha)\epsilon_1}]^{1/s} \leq 2e^{\epsilon_1}-2
	\end{multline*}
\end{itemize}
Taking $s=\alpha$, the inequality above holds. Thus, the first inequality holds so $b=\frac{\alpha}{\epsilon_1}$ is sufficient for differential privacy. 

Due to the space limit, we omit the details of second inequality. Because it is similar to the first part. So we can get the proof of privacy for $b=\frac{\alpha}{\epsilon_2}$ directly. 

\emph{\textbf{Theorem 4}:} The Re-DPoctor satisfies $w$-day $\epsilon$-differential privacy.

\emph{Proof:} According to Axiom 2.1.1 in \cite{Kifer2010Towards}, post-processing perturbed data maintain privacy as long as it does not use the sensitive information. Since among all the components, only the partition and perturbation process access to the raw data, while the others operate on the perturbed data.  Thus, if we can prove that these two mechanisms together satifsfies $w$-day $\epsilon$-differential privacy, the Re-DPoctor will satisfy $w$-day $\epsilon$-differential privacy.

According to Theorem 4, as previous proved, at  $i$, the partition process statisfies $q\cdot \epsilon_i$-differential privacy. According to Theorem 1, at  $i$, the perturbation process satisfies $(1-q)\cdot \epsilon_i$-differential privacy for applying Laplace noise. So for any  $i$, the Re-DPoctor provides $\epsilon_i$-differential privacy. Since the adaptive budget allocation mechanism guarantees for any sliding window $w$ that $\sum_{k=i-w+1}^{i}\epsilon_k\leq \epsilon$. Consequently, Re-DPoctor satisfies $w$-day privacy.

\section{Experimental Evaluation}

In this section, we evaluate the performance of Re-DPoctor on real health data. We have conducted real experiments on captured heart rates from wearable devices attached to a hospital patient during three months.

In the experiments, we set $\theta_P=0.8$, $\theta_I=0.2$, $\theta_D=0$ and $m=3$ for the PID controller. In Adaptive budget allocation, we set $\phi=0.2$.  In Partitioning, we use $T_D =30$, $T_R=15$, $T_L=4$ as the thresholds. Because heart rate usually changes between 50 and 200 and we track our w-day window as 14 days. So we define our sensitivity $\alpha = \frac{150}{14}$. Without explanation, we set $w=14$ and $\epsilon=3$ for all databases.

\begin{figure}[!t]
	\centerline{\includegraphics[width=8.9cm,clip]{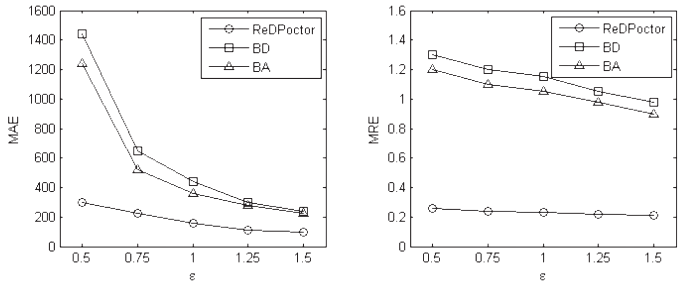}}	
	\caption{Utility comparision when $\epsilon$ changes ($w$=14)}		
\end{figure}

\begin{figure}[!t]
	\centerline{\includegraphics[width=8.9cm,clip]{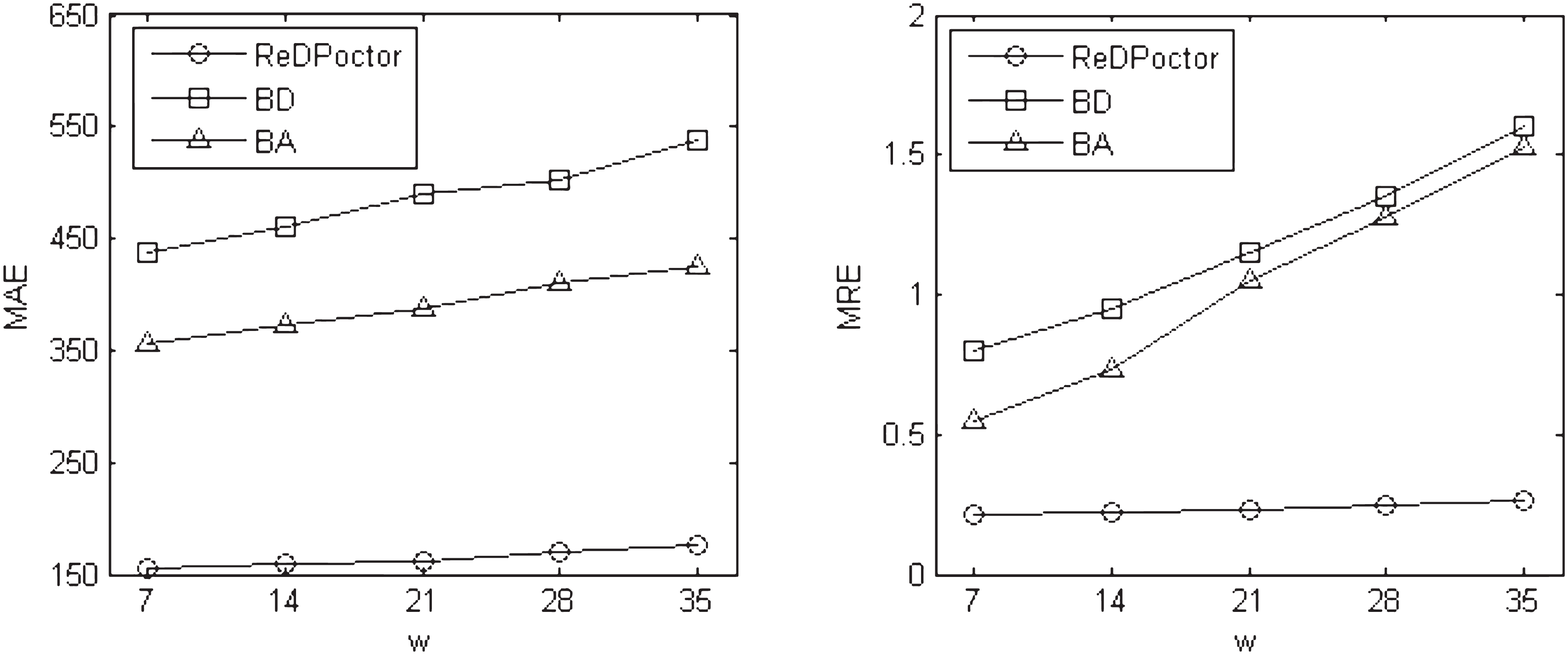}}	
	\caption{Utility comparision when $w$ changes ($\epsilon$=1)}		
\end{figure}

We use Mean Absolute Error(MAE) and Mean Relative Error( MRE) as the utility metrics to evaluate the performance of our scheme. The bound $\gamma$ is set to 0.05\% of $\sum_{i=1}^{n}x_i$ in order to mitigate the effect of extra small bins which could result from the take-off of the watch.

\textbf{Utility vs Privacy}:
Figure 2 investigates how MAE and MRE change with various $\epsilon$ values and makes the comparison between Re-DPoctor and BA and BD\cite{kellaris2014differentially}. We can see that with the increasing of $\epsilon$, both MAE and MRE of the dataset decrease. It is natural because a larger $\epsilon$ means smaller boise. Also, We can see that MAE and MRE both are smaller than BD and BA over the whole time period.

The better utility performance of Re-DPoctor contributes to three reasons. First, the Re-DPoctor adaptively adjust the sampling and allocate the privacy budget more appropriately. Within the fixed total budget, it samples the days with useful data and allocates more budget to them. Second. the Re-DPoctor has a more available budget for perturbation than other methods at any $w$ day window. In BD and BA, part of the budget is used for calculating the similarity. Third, the proper partition mechanism recognizes the patterns and improves the accuracy of released data.

\textbf{Utility vs $w$}:
In figure 3, we compare Re-DPoctor with BA and BD while varying $w$ values. We can see that the MAE and MRE of BD and BA increase greatly when $w$ increases. When $w$ increases, in order to ensure the total budget less than $\epsilon$, BA may skip the day which may contain useful data and results larger errors. In contrast, Re-DPoctor is more stable because it takes the window size and remaining budget into consideration and adaptively change the budget of next sampling point.

\textbf{Effect of Partitioning}: 
We also conduct two experiments of Re-DPoctor on the same dataset with and without partition to evaluate the effects of our partition mechanism. We can see from the results of Table 1 that the partition reduces MAE and MRE significatly. Therefore, we can conclude that partition can not only preserve the patterns but also improves the utility of released data.

\begin{table}
	\renewcommand{\arraystretch}{1.2}
	\caption{Utility with or without partition}
	\label{Utility with or without partition}
	\centering
	\begin{tabular}{|c||c|c|}
		\hline
		& With Partition & Without Partition\\
		\hline
		MAE & 156 & 355\\
		\hline
		MRE & 0.23 & 0.36\\
		\hline
	\end{tabular}
\end{table}

\section{Conclusions}
In this paper, we proposed Re-DPoctor, a real-time health data releasing scheme with $w$-day differential privacy achieving both utility and privacy guarantee. We designed a framework for Re-DPoctor consisting of mechanisms of adaptive sampling, adaptive budget distribution, partition, perturbation,ﬁltering, and feature extraction. The privacy analysis proves that Re-DPoctor satisﬁes $w$-day differential privacy. Experiments on real health data show that Re-DPoctor outperforms other methods and achieves both utility and privacy required.

\section*{Acknowledgement}
This work was supported by NSFC under grant 61402405 and Zhejiang Natural Science Foundation under grant No. LR16F020001.

\bibliographystyle{IEEEtran}
\bibliography{DP6pages}

% that's all folks
\end{document}